\documentclass[twocolumn,showpacs,preprintnumbers,amsmath,amssymb,prl,superscriptaddress]{revtex4}

\usepackage{graphicx}
\usepackage{color}

\usepackage{hyperref}

\begin{document}

\title{Spin-state transition in the Fe-pnictides}

\author{H.~Gretarsson}
\affiliation{Department of Physics, University of Toronto, 60
St.~George St., Toronto, Ontario, M5S 1A7, Canada}
\author{S. R.~Saha}
\author{T. Drye}
\author{J. Paglione}
\affiliation{Center for Nanophysics and Advanced Materials, Department of Physics, University of Maryland, College Park, Maryland 20742, USA}
\author{Jungho~Kim}
\author{D.~Casa}
\author{T.~Gog}
\affiliation{Advanced Photon Source, Argonne National Laboratory,
Argonne, Illinois 60439, USA}
\author{W.~Wu}
\author{S.~R.~Julian}
\affiliation{Department of Physics, University of Toronto, 60
St.~George St., Toronto, Ontario, M5S 1A7, Canada}
\author{Young-June~Kim}
\email{yjkim@physics.utoronto.ca} \affiliation{Department of
Physics, University of Toronto, 60 St.~George St., Toronto, Ontario,
M5S 1A7, Canada}

\date{\today}

\begin{abstract}

We report a Fe K$\beta$ x-ray emission spectroscopy study of local magnetic moments in the rare-earth doped iron pnictide Ca$_{1-x}RE_x$Fe$_2$As$_2$ ($RE$=La, Pr, and Nd). In all samples studied the size of the Fe local moment is found to decrease significantly with temperature and goes from $\sim \!\! 0.9 \mu_B$ at T = 300 K to  $\sim \!\! 0.45 \mu_B$ at T = 70 K.
In the collapsed tetragonal (cT) phase of Nd- and Pr-doped samples (T$<$70K) the local moment is quenched, while the moment remains unchanged for the La-doped sample, which does not show lattice collapse. Our results show that Ca$_{1-x}RE_x$Fe$_2$As$_2$ ($RE$= Pr and Nd) exhibits a spin-state transition and provide direct evidence for a non-magnetic Fe$^{2+}$ ion  in the cT-phase, as predicted by Yildirim. We argue that the gradual change of the the spin-state over a wide temperature range reveals the importance of  multiorbital physics, in particular the competition between the crystal field split Fe 3$d$ orbitals and the Hund's rule coupling.

\end{abstract}



\maketitle

The interesting orbital physics found in many 3$d$ and 4$d$  transition metal compounds, such as manganites \cite{Nagaosa2000,Hotta2006} and ruthenates \cite{Hotta2006}, seems to play an important role in the iron based superconductors as well \cite{Kruger2009,Lv2009,Lv2010,Lee2009,Lee2010,Chan2009,Chan2010}.
In the iron pnictides, many low-energy probes such as transport \cite{Fisher2010}, scanning tunnelling microscopy  \cite{Davis2010}, inelastic neutron scattering  \cite{Pengcheng2009}, angle-resolved photoemission spectroscopy \cite{Shen2011, Shimojima2010}, and most recently magnetic torque measurements \cite{Kasahara2012}  have reported a strong in-plane anisotropy of electronic properties. These results have spurred a great deal of interest in the orbital physics of the iron pnictides, in particular the possibility of orbital order
\cite{Kruger2009,Lv2009,Lv2010,Lee2009,Lee2010,Chan2009,Chan2010}.

An important aspect of the orbital physics is the competition between the Hund's rule coupling constant ${\emph J_H}$ and the crystal field splitting, $\Delta_{{\rm CF}}$. In the case of LaCoO$_3$, the energy scales of $\Delta_{{\rm CF}}$ and ${\emph J_H}$ are similar, resulting in spin-state transition; Co$^{3+}$ ions take on a low-spin state (S=0) at low temperature, but go into thermally excited high/intermediate-spin (S=2 or S=1) states at elevated temperature \cite{Blasse1965,Sawatzky1996}. Applied pressure, either chemical or hydrostatic, can alter $\Delta_{{\rm CF}}$ through lattice changes and affect the balance between ${\emph J_H}$ and $\Delta_{{\rm CF}}$. A high-spin state  was for instance observed in the case of La$_{0.82}$Sr$_{0.18}$CoO$_3$ in which Sr dopants exert negative chemical pressure \cite{Lengsdorf2007}, while a low-spin state was observed under hydrostatic pressure \cite{VankoPRB2006,Lengsdorf2007}.

Among the iron based superconductors, CaFe$_2$As$_2$ offers perhaps the best system to investigate the competition between $\Delta_{{\rm CF}}$ \cite{CF} and ${\emph J_H}$, and its  effect on the spin-state. Like many iron pnictides, CaFe$_2$As$_2$ goes from a high temperature tetragonal phase (T-phase) to an orthorhombic and antiferromagnetically (AFM) ordered phase, below $T_N \approx$ 170 K \cite{Canfield2008}. More importantly, CaFe$_2$As$_2$ takes on yet another structural phase at low temperatures through application of a modest pressure of 0.35 GPa \cite{Kreyssig2008} or chemical doping, with rare-earths \cite{Saha2012} or phosphorus \cite{Kasahara2011}. Upon entering this phase, known as the collapsed tetragonal phase (cT-phase), the lattice undergoes a $\sim\!\! 10 \%$ reduction along the {\em c}-axis  and an $\sim\!\! 2 \%$ increase along the {\em a}-axis. This is accompanied by a disappearance of the AFM order \cite{Kreyssig2008}, supression of spin fluctuations  \cite{Pratt2009}, and recovery of Fermi liquid behavior \cite{Kasahara2011}.  It is thus clear that an unusually dramatic lattice instability exists in CaFe$_2$As$_2$ and its doped counterparts, and that the magnetic and the electronic structure are strongly influenced by this instability. In particular, the change in $\Delta_{{\rm CF}}$, due to the distortion of the FeAs$_4$ tetrahedra in the cT-phase \cite{Saha2012}, can result in a very different spin-state of the Fe$^{2+}$ ion.

\begin{figure*}
\advance\leftskip-2cm
\advance\rightskip-1cm
\includegraphics[width=1.8\columnwidth]{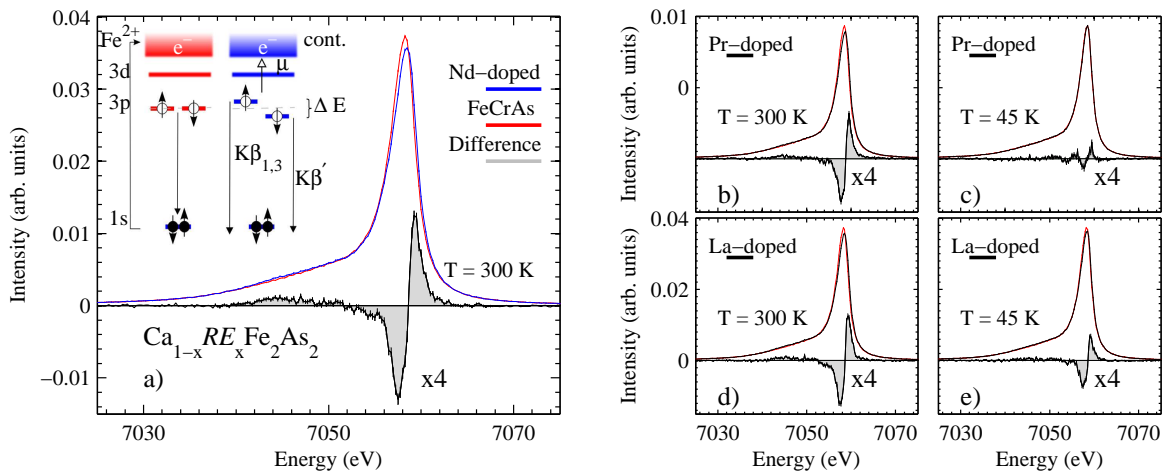}
\caption{(Color online)
(a) Comparison of the K$\beta$ emission line for Nd-doped sample and FeCrAs taken at room temperature. The difference spectrum (grey) was magnified by a factor of 4 here and the rest of the figure.The inset shows the K$\beta$ emission process in the atomic limit (see text). In (b)-(e) the temperature dependence of this difference, obtained in the same way as in (a), is shown for  Pr-doped and La-doped samples.} \label{fig01}
\end{figure*}

In this Letter, we use x-ray emission spectroscopy (XES), which is a very sensitive probe of local instantaneous spin moment, to investigate the spin-state of the Fe$^{2+}$ ion in Ca$_{0.92}$Nd$_{0.08}$Fe$_2$As$_2$ (Nd-doped), Ca$_{0.85}$Pr$_{0.15}$Fe$_2$As$_2$ (Pr-doped), and Ca$_{0.78}$La$_{0.22}$Fe$_2$As$_2$ (La-doped) as a function of temperature.
We find that local moments disappear in the cT-phase of the Nd- and Pr-doped samples, indicating that the Fe$^{2+}$ ions go through a spin-state transition by taking on the low-spin state; this confirms earlier calculations by Yildirim \cite{Yildirim2009}. X-ray absorption near edge spectra (XANES) taken at the Fe $K$-edge reveal large changes in the electronic structure due to the collapse of the lattice. Surprisingly for all the samples studied, the Fe local moment is found to decrease significantly with decreasing temperature in the T-phase, from $\sim \!\! 0.9 \mu_B$ at T = 300 K to  $\sim \!\! 0.45 \mu_B$ below T$\sim$70 K. This behavior could be described as a thermally induced spin-state crossover which arises due to the competition between $\Delta_{{\rm CF}}$ and ${\emph J_H}$. Our findings seem to suggest that spin-state degeneracy could be important for the understanding of iron pnictides, as recently proposed by Chaloupka and Khaliullin \cite{Khaliullin2012}.

The XES measurement was performed at the Advanced
Photon Source on the undulator beamline 9ID-B using the identical setup as in Ref.~\cite{emission2011}. The XANES spectra in the partial fluorescence yield mode (PFY-XANES) was measured by monitoring the Fe K$\beta$ emission line across the Fe $K$-edge.  X-ray diffraction measurements were performed using a Cu tube source with a graphite (002)  monochromator, and a four-circle diffractometer. For all temperature dependence studies closed-cycle refrigerators were used.
Details of the growths and characterization of the single-crystal samples have been reported in earlier publications \cite{Wu2011,Saha2012}.


The local moment sensitivity of the Fe K$\beta$ emission line $(3p
\rightarrow 1s)$ originates from a large overlap between the $3p$ and
$3d$ orbitals. This interaction is mainly driven by the presence of a net magnetic moment $(\mu)$ in the $3d$ valence shell  \cite{Tsutsumi1976,Peng1994} and causes the  K$\beta$ emission line to split into a main peak K$\beta_{1,3}$ and a low-energy satellite K$\beta^\prime$. A schematic diagram of the Fe K$\beta$ emission process is  shown in Fig. \ref{fig01} (a) inset for both non-magnetic (red) and magnetic (blue) Fe$^{2+}$ in the atomic limit. Filled and empty circles represent electrons and holes, respectively, and $\Delta E$ represents the splitting of  K$\beta_{1,3}$ and K$\beta^\prime$. Information on the local moment of Fe can be extracted using the overall shape of the Fe K$\beta$ emission spectra by applying the  integrated absolute difference (IAD) analysis \cite{Vanko2006}. In Fig. \ref{fig01} (a) we demonstrate how this method works by showing Fe K$\beta$ XES data for  the Nd-doped sample  taken at T = 300 K along with a non-magnetic FeCrAs reference spectrum \cite{emission2011,Wu2009,Ishida1996}. Relative to the main line in FeCrAs, we see that the Nd-doped K$\beta_{1,3}$ peak shifts towards higher energy, while the intensity and the width of this peak also change; a contribution from K$\beta^\prime$ on the lower energy side becomes visible now.  These changes are all attributed to the existence of a local moment. To follow the IAD procedure from Ref.~\citenum{Vanko2006}, the area underneath each spectrum was normalized to unity. The reference spectrum was then subtracted from the sample spectrum, and the resulting difference plotted. For display purpose, the difference was magnified by a factor of 4. The IAD value can be extracted by integrating the absolute value of the difference spectrum. This quantity is found to be linearly proportional to the local spin magnetic moment of the Fe atom \cite{Vanko2006}. This method has recently been applied to study various iron-based superconductors \cite{emission2011, Chen2011, Simonelli2012}.

Fe K$\beta$ emission lines obtained at different temperatures are shown in Fig. \ref{fig01} (b) - (e) for both the Pr- and La-doped samples. At T = 300 K the samples show the same characteristics as the Nd-doped. However, at T = 45 K significant changes can be observed, the K$\beta_{1,3}$ shifts towards lower energy and the contribution from K$\beta^\prime$ is supressed. This is well captured in the difference spectra and provides evidence for a decreased local moment. The change is much larger for the Pr- than for the La-doped sample; in fact a complete supression of the difference spectra is observed for the Pr-doped sample. It should be noted that such a strong thermally induced change is surprising given that neither the presence of long-range order nor carrier doping had any affect on local magnetic moment in other iron based superconductors \cite{emission2011}.

\begin{figure}
\includegraphics[width=\columnwidth]{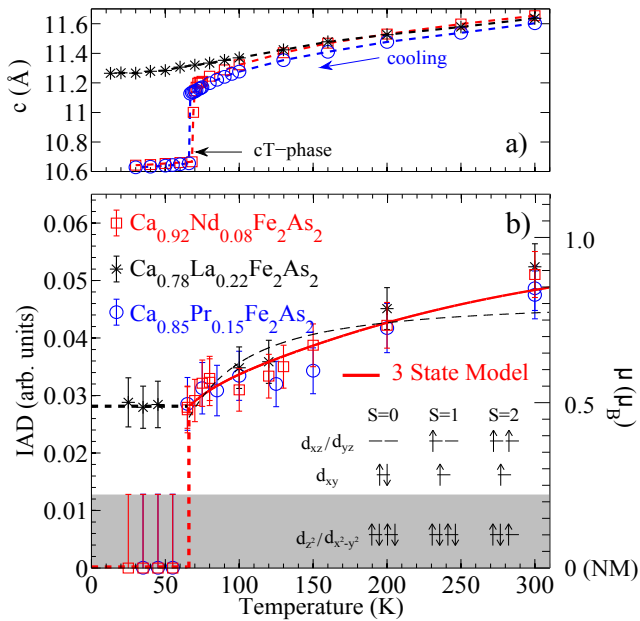}
\caption{(Color online) The temperature dependence of (a) c-axis lattice parameters, and (b) the IAD values derived from the XES spectra. Same symbols are used in both panels. On the right hand side of panel (b), the local magnetic moment scale ($\mu$) as described in Ref. \citenum{emission2011} are shown. Thick dashed lines are guides to the eyes. The solid red line and thin dashed black lines are fits to 3-state and 2-state models, respectively. The spin and orbital configuration of the 3-state model used for the fitting is shown in the inset. The grey area represents the detection limit of the IAD method. } \label{fig02}
\end{figure}

In order to extract quantitative information about the evolution of the local moment in these samples we have studied detailed temperature dependence of the IAD values.  The results are plotted in Fig. \ref{fig02} (b), in which the right-hand side of the figure is the local moment scale determined from the IAD value of $\rm K_2Fe_4Se_5$ \cite{emission2011}. Dashed lines are given as guides to the eye. At room temperature all three samples have local moments around $\sim\!\! 0.9{\rm \mu_B}$. With decreasing temperature this local moment decreases and at T = 70 K the local moment has already been reduced by a factor of 2, to a value of $\sim\!\! 0.45{\rm \mu_B}$. Below T = 70 K a noticable difference among the three samples can be observed. Both Nd- and Pr-doped samples show an abrupt drop in the IAD value down to an undetectable level, while the La-doped sample shows no change in the same temperature range.  The detection limit of the IAD technique is shown as a grey area and indicates that at low temperature the local moment for Nd- and Pr-doped samples is $<\!\!0.2{\rm \mu_B}$. The evolution of the c-axis lattice constant, as determined from the position of the (008) Bragg peak, is shown in Fig. \ref{fig02} (a).  Data were obtained on cooling and shows the T- to cT-phase transition clearly for the Nd- and Pr-doped samples. One should note that the lattice constant change is unusually large even well above the cT-T transition temperature. The linear thermal expansion coefficient of the La-doped sample is almost a factor of 10 larger than that of Ba(Fe,Co)$_2$As$_2$ \cite{Luz2009}.

The change of the local moment from T= 300 K to 70~K is puzzling.
In LaCoO$_3$,  a continuous increase in the local spin moment as a function of increased temperature was observed \cite{VankoPRB2006}. Using a three spin-state model, this behavior was explained as arising from spin-state transition, in which the low-spin Co$^{3+}$ ions ($S$=0) are  thermally excited into a magnetic state ($S$=1 or 2), resulting in a larger local moment. To determine whether changes in the spin-state can account for the observed change in local moment, we follow the analysis on LaCoO$_3$ \cite{VankoPRB2006} and analyze the data in terms of thermally excited localized three spin-states model \cite{Saitoh1997}. The effective moment can then be expressed as:

\begin{equation}
\overline{\mu}(T)=A\sum_{i=0}^{2} g\mu_B\sqrt{S_i(S_i+1)}\nu_ie^{-\Delta_i/k_BT}/Z
\end{equation}

\noindent where $i$ indexes the states with spin $S_i$ and energy $\Delta_i$ relative to the non-magnetic spin state ($S_0$=0), $\nu_i$ is the degeneracy factor and $g$=2, while $Z=\sum_{i=0}^{2}\nu_ie^{-\Delta_i/k_BT}$. We assume that the partially itinerant nature of the iron pnictides reduces the local moment size by an overall scaling factor $A$ \cite{emission2011}. We include both intermediate ($S_1$=1) and high spin-state ($S_2$=2) of the Fe$^{2+}$ ($d^6$) ion. In the ionic picture the FeAs$_4$ tetrahedral crystal field splits the 3$d$ orbitals into lower $e$ and upper $t_{2}$; a small tetragonal crystal field further splits the $t_{2}$ states into upper $d_{xz}/d_{yz}$ and lower $d_{xy}$ \cite{Kruger2009}. This gives us an estimate for the orbital degeneracy and results in an overall degeneracy (including spin) of  $\nu_0$=1, $\nu_1$=6 and $\nu_2=10$ (see inset of Fig. \ref{fig02} (b)). This spin-state crossover model describes the temperature dependence of the local moments above 70 K very well, as shown as the solid line in Fig. \ref{fig02} (b), which was obtained using  $\Delta_1 = 8$ meV, $\Delta_2 = 45$ meV and an overall scaling factor of $A = 0.3$. The model also reveals that the continuous drop in local moment comes from high-spin Fe$^{2+}$ ions going into a low-spin state, i.e. the ratio of $S_0$:$S_1$:$S_2$ spin-states goes from $\sim$0.15:0.60:0.25 at 300 K to $\sim$0.40:0.60:0 at 70 K. We also tried to fit the data using just two spin states (S=0 and S=2). The fit result plotted in Fig. 2(b) quickly reaches a plateau, underestimating the local moment at higher temperatures.  We note that $\Delta_i$ is a phenomenological energy splitting between spin-states, and does not correspond to real energy scales such as  $\Delta_{{\rm CF}}$ or ${\emph J_H}$. The behavior of the La-doped sample at lower temperatures (T$<$70K) deviates from the model calculation, which could originate from our assumption that $\Delta_i$ and $\nu_i$ are constants, when in principle they can vary with temperature.

Our analysis suggests that the Fe$^{2+}$ ions consists of a mixture of different spin-states; this provides an important clue as to the energy scale of
 $\Delta_{{\rm CF}}$ and ${\emph J_H}$. That is, the energy scale of $\Delta_{{\rm CF}}$ and ${\emph J_H}$ in Ca$_{1-x}RE_x$Fe$_2$As$_2$ must be comparable in order for the spin-state crossover to occur. Since a high-spin Fe$^{2+}$ ion has a larger ionic radius than its low-spin counterpart \cite{Shannon1976} the lattice is forced to expand as the local moment increases. This could help explain the large linear thermal expansion coefficient observed in all our samples \cite{Radaelli2002}. We emphasize that Ca$_{1-x}$RE$_x$Fe$_2$As$_2$ might be a rare example where a spin-state crossover can be observed. We speculate that the small ionic radius of Ca, compared to Sr or Ba, exerts negative chemical pressure in Ca$_{1-x}RE_x$Fe$_2$As$_2$ and splits the spin states. In (Ba,Sr)Fe$_2$As$_2$ these spin states might be degenerate, resulting in a temperature independent local moment. In a similar experiment on SrFe$_2$As$_2$, indeed no sign of spin-state crossover was observed \cite{Bondino2012}. The idea of degenerate Fe$^{2+}$ spin-states has recently been proposed in order to understand the magnetic behavior of the iron pnictides \cite{Khaliullin2012}.

\begin{figure}
\includegraphics[width=\columnwidth]{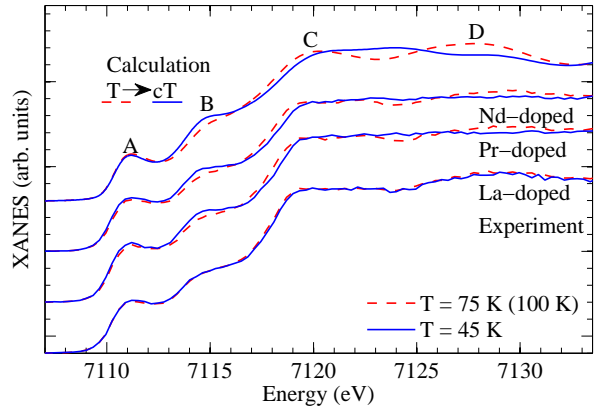}
\caption{(Color online) Fe K-edge x-ray absorption near edge spectra taken in the partial fluorescence yield mode by monitoring the Fe K$\beta$ emission line. Spectra for  Nd- and Pr-doped samples were taken at T = 75 K and 45 K, and at T = 100 K and 45 K for La-doped.  The simulated spectrum was calculated for a Nd-doped crystal in the T and cT phase. It has been shifted in energy to match the pre-edge peak A for the experimental scans. Spectra were offset for clarity.} \label{fig03}
\end{figure}

The sudden collapse of the Fe$^{2+}$ moments below T = 70 K for both the Pr- and Nd-doped samples is too sharp to be explained by the crossover model in Eq. (1). Large structural changes can often be associated with changes in the electronic structure and thus affect the magnetism directly \cite{Chen2011}.  To investigate whether that is the case in Fig. \ref{fig03} we show the XANES spectra obtained above and below the cT-transition temperature. The absorption spectra consist of a prominent pre-edge peak A and three higher energy features B, C, and D. Similar features have been observed in BaFe$_2$As$_2$ \cite{Bittar2011} and were found to be independent of doping. Now both Nd- and Pr-doped samples show a difference going from T = 75 K to T = 45 K. In particular C shifts towards higher energy and B becomes more pronounced, while A stays more or less unchanged. The La-doped sample showed no noticable difference going from T = 100 K to T =45 K.  In order to see whether the T- to cT-phase transition could explain this difference we simulated XANES spectra using the FDMNES code \cite{FDMNES} with a radius of 8${\rm \AA}$.  The reported tetragonal crystal structure of Ca$_{0.91}$Nd$_{0.09}$Fe$_2$As$_2$ at T = 105 K and T = 80K \cite{Saha2012} was used for T- and cT-phase, respectively. 
The simulated spectra capture fairly well the changes we observed for the Nd- and Pr-doped samples.

Theoretical calculations show that the  T- to cT-phase transition is accompanied by the formation of interlayer As-As dimers \cite{Yildirim2009} which are believed to form below the critical interlayer As-As distance of 3${\rm \AA}$ and cause the lattice to suddenly collapse \cite{Saha2012}. This collapse
changes both the As-Fe-As angle and the Fe-As bond length, distorting the FeAs$_4$ tetrahedra. In previous XANES studies of Fe-pnictides, features similar to C and D have been assigned to hybridized Fe/As 4p states \cite{Bittar2011, Ignace2010}. These states are also known to vary strongly due to local bonding and symmetry effects \cite{Ignace2010}. The loss of local magnetic moment can  thus be understood as coming from the distortion of the FeAs$_4$ tetrahedra, which changes the electronic structure and forces the Fe$^{2+}$ ion to the low spin-state. In a recent NMR experiment (much slower probe) \cite{Long2012} evidence for a large suppression of the local Fe moment in the cT-phase have been reported, supporting our findings.

In summary, we have studied the spin-state of the Fe$^{2+}$ ion in Ca$_{1-x}$RE$_x$Fe$_2$As$_2$ (RE=La,Pr, and Nd) as a function of temperature. The continuous decrease of the local moment in the T-phase of all the samples is explained through spin-state crossover which cause the local moment to decrease by a factor of two. The ionic radius of the Fe$^{2+}$ ion decreases with  decreasing moment size, allowing the  $c$-axis to contract,  which can explain the large linear thermal expansion coefficient in these compounds. In the Nd- and Pr-doped samples this reduction causes the interlayer As-As distance to cross the critical value of 3${\rm \AA}$,  resulting in the formation of As-As dimers and transition into the collapsed tetragonal phase, which is accompanied by a total loss of local moment as predicted by Yildirim \cite{Yildirim2009}.

\acknowledgements{We would like to thank David Hawthorn for initiating experiments on these samples. Research at the University of Toronto was supported by the NSERC, CFI, OMRI, and CIfAR. Research at the University of Maryland was supported by AFOSR-MURI Grant No. FA9550-09-1-0603 and
NSF-CAREER Grant No. DMR-0952716. Use of the APS was supported by the U.S. Department of Energy, Office of Science, Office of Basic Energy Sciences, under Contract No. W-31-109-ENG-38.  }

\end{document}